\documentclass[prl,twocolumn,amsmath,amssymb,floatfix, showkeys]{revtex4}
\usepackage{graphicx}% Include figure files
\usepackage{bm}% bold mat
\usepackage{amssymb}
\usepackage{color}
\usepackage{epsfig}
\usepackage{subfigure}
\usepackage{graphicx}% Include fgure files
\usepackage{epsfig}
\usepackage{subfigure}
\usepackage{amssymb,amsfonts,amsmath}

\newcommand{\be}{\begin{equation}}
\newcommand{\ee}{\end{equation}}
\begin{document}

\title {Dislocation-mediated growth of bacterial cell walls}

%Date: \today

%\author{  Ariel Amir}$^{1}$, Stefano Borini$^{2}$, Yuval Oreg$^{1}$, Yoseph Imry$^{1}$}
\author{Ariel Amir and David R. Nelson}
\affiliation { Department of Physics, Harvard University, Cambridge, MA 02138, USA}

\begin{abstract}
Recent experiments have illuminated a remarkable growth
mechanism of rod-shaped bacteria: proteins associated with cell
wall extension move at constant velocity in circles oriented
approximately along the cell circumference (Garner et al., \emph{Science} (2011), Dom\'inguez-Escobar et al. \emph{Science} (2011), van Teeffelen et al. \emph{PNAS} (2011). We view these as dislocations in the partially ordered peptidoglycan
structure, activated by glycan strand extension machinery, and study theoretically the dynamics of these interacting
defects on the surface of a cylinder. Generation and motion of these interacting defects lead to surprising effects arising from the cylindrical geometry,
with important implications for growth. We also discuss how long range elastic interactions and turgor pressure affect
the dynamics of the fraction of actively moving dislocations in the bacterial cell wall.
 \end{abstract}

%% When adding keywords, separate each term with a straight line: |

%% Optional for entering abbreviations, separate the abbreviation from
%% its definition with a comma, separate each pair with a semicolon:
%% for example:
%% \abbreviations{SAM, self-assembled monolayer; OTS,
%% octadecyltrichlorosilane}

% \abbreviations{}
 \maketitle

Bacterial cell walls are composed of peptidoglycan (also called murein), which endows them with shape and rigidity. Their architecture and growth have been the subject of active research for many decades, in particular for gram-negative bacteria whose cell walls consist of single or few layers of glycan strands crosslinked by peptides \cite{young, sun, Scheffers, vollmer2}. While some models assume a very ordered structure, recent experimental work \cite{jensen} suggests the structure is more disordered. We view the peptidoglycan mesh as a partially ordered two-dimensional crystal  with a large number of defects to account for the disorder in the structure. The rod shape of many bacteria (e.g.: \emph{Escherichia coli}), together with mutant variants that grow but fail to complete cell division, make a cylindrical geometry a natural one to study. To easily add material to this ordered structure, one must clearly break the periodicity and create a defect in the structure. An especially important class of defects are termed dislocations, known to be important in determining the mechanical properties of metals and other crystalline or polycrystalline materials, such as their strength and plasticity \cite{hirth}. Dislocations are known to have long-ranged elastic interactions, which are logarithmic in the distance (but not isotropic, as is the case for vortices in superfluids). The ``elementary charge" of these topological defects is the Burgers vector $\vec{b}$, which is often at right angles to the direction of insertion of the new strand of material, and is a lattice vector of the structure. Here, we apply dislocation theory to the problem of bacterial growth, proposing a simplified model inspired by the elongation of bacterial cell walls, which we are able to solve both analytically and via computer simulations. Fig. 1 illustrates an idealized picture of defects in a cylindrical geometry. For simplicity, we show a square lattice with lattice vectors parallel and perpendicular to the cylinder's long axis., although the actual peptidoglycan mesh is rectangular (and in fact the lattice vectors might have a non-trivial angle with respect to the cylinder's axis \cite{wang2012}).

\begin{figure}[b!]
\includegraphics[width=0.5\textwidth]{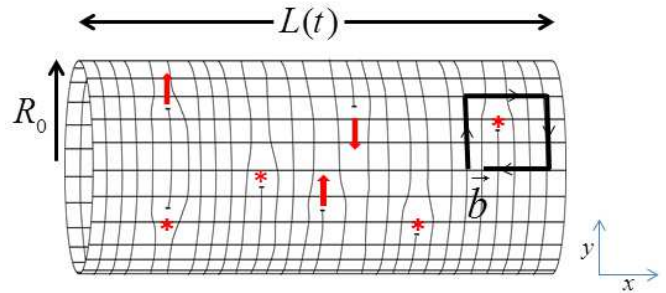}
\caption {Schematic illustration of active (arrows) and inactive (asterisk) dislocations in an otherwise ordered peptidoglycan mesh. The dislocations with arrows attached are activated by the enzymatic machinery and move with velocity $v$. Those with asterisks are inactive. }
\label {dislocation}
\end{figure}

The model treated below will in fact consist of a \emph{large} number of dislocations, and so the structure we treat is far from a perfect crystal. Working with defects of a crystal provides a convenient and numerically efficient method to take the disorder into account. We expect that the unit cell of Fig. 1 typically contains two glycan strands, as it is only this larger unit cell that respects the local crystalline symmetry; see for example Refs. \cite{vollmer,nelson_review}. Recent experiments on both gram-negative \cite{shaevitz} and gram-positive bacteria \cite{garner,escobar} track fluorescently labeled proteins known to correlate strongly with the addition of peptidoglycan subunits, and have shown that these proteins move at roughly constant velocity, approximately along the cylinder's circumference. We view these strand extension centers as edge dislocations in the ordered structure, with a Burgers vector oriented along the cylinder's long axis (the direction of the Burgers vector depends on the direction of insertion of the new strand, see Fig. 1). Extending the end of an inserted strand (\emph{i.e.}, the core of an edge dislocation) involves breaking peptide bonds to allow extra sugar units into the lattice, together with additional short peptide crosslinks \cite{young}. In dislocation theory, this type of motion is referred to  as dislocation \emph{climb}. In the following, we treat the protein motion and the dislocation motion synonymously, assuming that the motion of the  MreB and its associated enzymes is fully correlated with the insertion of new material into the cell wall. This idea was introduced in \cite{nelson_review}, and here we develop it further and deduce various biological insights and predictions. As pointed out by Burman and Park \cite{burman}, glycan strand extension is somewhat analogous to the action of a DNA polymerase. Our dislocation perspective allows us to take into account long range elastic interactions between the ``murein extension centers" of Ref. \cite{burman}.

Although extensive work has been done on dislocation theory over the last century \cite{hirth}, this biophysics problem is quite different from materials science and condensed matter physics in several respects: the climb of dislocations necessarily involves the exchange of material, which in 3d crystals involves diffusion of interstitials or vacancies. Hence, except at high temperatures, dislocation glide dominates the dynamics \cite{glide_footnote}. Here, dislocation climb is the central process mediating cell wall growth, with sugars and amino acids essential for the climb process synthesized and arriving from the interior of the bacteria. A second, obvious, difference, regards the unusual cylindrical geometry which we study here.  %as far as we know, the interactions of dislocations in cylindrical geometry has not been studied %previously.
This feature leads to a number of interesting properties, such as the exponential decay of dislocation interactions along the cylinder's long axis, as discussed below. Dislocations in a cylindrical geometry were considered in a very different biophysics problem in Ref. \cite{hartman}, studying tail-sheath contraction in a bacteriophage. A final difference from conventional materials science is that here the dislocation climb itself alters the geometry, as it is this very motion that grows the cell wall along the axis of the cylinder. This idea can be used to estimate the number of actively climbing dislocations per bacterium, using experimental data:  The typical time of division of the bacteria is of the order of tens of minutes, in which the bacteria elongate by approximately 1 $\mu m$. The lattice spacing of the peptidoglycan along the long axis, $b$, is believed to be of the order of several nanometers \cite{young}.
If we assume the area of the hemispherical end caps remains fixed, the dynamics of the cylindrical surface area $S(t)$ (taking all dislocations as completely independent, for now), is given by:
\be \frac{dS}{dt}=N_{ac} v b, \label{growth} \ee where $N_{ac}$ is the total number of actively moving dislocations and $v$ is their velocity.
Taking the measured velocity $v$ to be several tens of nanometers/second \cite{shaevitz,garner,escobar}, we find that \emph{a few tens} of active dislocations moving along the circumference and growing the cell wall would be sufficient to account for the measured growth rate. This estimate is consistent with pioneering work of Ref. \cite{burman}, obtained using a very different method of radioactive labeling. Current technology does not yet enable a direct determination of $N_{ac}$, since only a subset of the total number of active dislocations is fluorescently marked \cite{shaevitz,garner,escobar}.

 The structure of the manuscript is as follows. We first define the model and the approximations made, and describe the force on a dislocation, including the elastic interactions between dislocations on a cylinder. The implications of the large turgor pressure inside the bacteria and the effects of the long ranged elastic interactions are highlighted. We then propose a novel set of equations for the growth dynamics and its coupling to the numbers of active and inactive dislocations, and show that they often lead to exponential lengthening of a single bacterium. The exponential growth rate itself depends on a few simple parameters, with a well defined microscopic interpretation. The theoretical expectations are illustrated and visualized by numerical simulations, both in the main text and the Supplementary Material.  A numerical calculation presented in  the Supplementary Material is used to estimate the disorder strength due to the elastic interactions with a large number of randomly positioned dislocations, chosen according to the biological parameters. Certain parameters of the rate-equations model (Eqs. (\ref{growth1}) and (\ref{growth2})), can not be determined from a numerical simulations, and are dictated by the underlying biochemistry, for example v(G) (the dependence of the dislocation velocity on the driving force) and the dislocation processivity $\gamma_1$.

\section {The model}
Interesting recent works have modeled and simulated the peptidoglycan structure in molecular detail, and have predicted a number of non-trivial phenomena which were indeed observed experimentally \cite{wingreen1, wingreen2}. Our approach is somewhat different (and perhaps complementary), since we study a highly simplified model and focus on a relatively dilute concentration of defects in an otherwise regular peptidoglycan mesh. As a result both analytic and numerical progress is possible, although we suppress fine details of the structure.

Consider, motivated by bacterial cell wall growth, a number of interacting edge dislocations with a Burgers vector $\vec{b}$ oriented in both directions along the cylinder's long axis, as illustrated schematically in Fig. 1. The radius of the cylinder, $R$, will be assumed to be constant, as is approximately true for rod-shaped bacteria. For simplicity, consider an infinite cylinder, neglecting the effects of the boundaries (Experiments indicate that strand extension dynamics does not change significantly near the cylinder's end caps, consistent with this assumption \cite{garner1}).
%For now, we neglect additional imperfections in the peptidoglycan mesh and focus on actively moving dislocations in an otherwise perfect lattice. The effect of frozen disorder on the dislocation dynamics \cite{nelson_review} will be treated in more detail elsewhere \cite{amir_nelson_PRE}.

We also simplify by assuming linear elasticity, except near dislocation cores. Various works have indicated, both from the theoretical \cite{boulbitch} and experimental \cite{shaevitz2} perspective, that the high osmotic pressure \cite{shaevitz2,whatmore}, referred to as turgor pressure in bacteria and plant cells, produces large strains compared to the relaxed state. We shall take the elastic moduli as constants, which in practice should be considered as effective constants relative to the working turgor pressure.
%
%Although extensive work has been done on dislocation theory over the last century, the physics problem naturally posed by the biological scenario is rather unique, in several respects. First of all, as explained before, here the \emph{climb} of the dislocation forms a substantial and relevant part of the dynamics, which is certainly not the case in condensed matter physics. The essential reason for it here is that the "building materials" which are essential for the climb process come from the third dimensions (the bulk of the bacteria), which has no analogue in three dimensional metals, for example. This point will also have an implication on the form of the force on the dislocations, to be shortly discussed. A second, obvious, difference, regards the peculiar geometry which we study here: as far as we know, the interactions of dislocations in cylindrical geometry has not been studied previously. This feature turns out to give rise to a number of interesting properties, such as  the exponential decay of the interactions along the cylinder's long axis, to be elaborated upon later.

In order for a dislocation to be able to ``climb" (\emph{i.e.}, move perpendicular to its Burgers vector \cite{hirth} and contribute to the growth process), various proteins have to be present at the dislocation core, responsible for ``recruiting" and assembling the sugars and peptides necessary for the construction of the peptidoglycan mesh. Hence, we separate the dislocations into two populations, active and inactive ones. Inactive defects still exert elastic forces on the active ones, and create an effective disordered energy landscape for them. They also represent favorable locations for the creation of new active dislocations: the elongation machinery can attach to the free strands (\emph{i.e.}, the inactive dislocations), creating active ones. The finite processivity of the elongation machinery will also give rise to the opposite process, whereby an active dislocation can become inactive when the machinery falls off. In general, the rates of these processes will not balance, since the system is never in steady-state but is instead constantly growing via dislocation climb, as it incorporates material from the third dimensions, a situation rarely encountered in conventional materials science \cite{hirth}.

\subsection {The force on a dislocation\label{force_section}}
 Elastic stresses exert a Peach-Koehler force \cite{hirth,peach} on a dislocation (analogous to the Magnus force acting on vortices \cite{lamb}) which in our case is given by \cite{weertman_comment}:

\begin{equation}
F_x= b \sigma_{xy}; \hspace {0.5 cm}
F_y= -b \sigma_{xx},\label{fy}
\end{equation}
 where $\sigma$ is the two-dimensional stress tensor of the peptidoglycan mesh, and we assume a Burgers vector along the x-axis.

%\begin{equation}
%F_i= b_k \sigma_{kj}\epsilon_{ij},\label{force}
%\end{equation}
%
%where $\epsilon_{ij}$ is the 2X2 antisymmetric unit tensor and $\sigma_{kj}$ the two-dimensional stress tensor of the peptidoglycan mesh. %\cite{shaevitz2}.
%
% In \cite{weertman}, a correction to the Peach-Koehler force is discussed, which is derived from energy considerations, and eliminates the excess
%force on a dislocation due to a hydrostatic pressure. The forces are still given by the above equations, but in three dimensions the stress
%tensor should be replaced by $\tilde{\sigma}=\sigma-\frac{1}{3}{\rm {tr}[\sigma],}$
% and in two dimensions one should replace $\frac{1}{3}\rightarrow\frac{1}{2}$.
%
%Essentially, the additional term is due to the contraction or expansion of the crystal as a whole which is associated with the climb of dislocations. However, in the case discuss here, we can safely assume that the additional material added in the growth process does not change the density of the two-dimensional crystal, since it comes from the bulk. Therefore, following the same reasoning as of Ref. \cite{weertman}, we conclude that it is Eq. (\ref{force}) which determines the force on a dislocation, also for climb processes.

For a cylinder of radius $R$ we have $\sigma_{yy}=2\sigma_{xx}= p R$ , where $x$ denotes the coordinate along the long axis, and $ p$ is the turgor pressure, and $F_y=-\frac{1}{2} b p R$. In addition to the contribution of the turgor pressure, if the free energy is changed by $U$ by the biochemical process of adding one unit cell to the peptidglycan mesh, it will contribute an additional force of $U/b$ in the $y$ direction, as can be seen using the principle of virtual work.

Under physiological conditions, we expect that the dislocations are in the overdamped regime, with the dislocation velocity proportional to the force, \emph{i.e.}, $v_i = \sum_j \mu_{ij} F_j$, where $\mu_{ij}$ is a mobility tensor with glide and climb components that depends on the extension machinery and the abundance of sugars, peptides, \emph{ etc.} In the following we assume that $\mu_{ij}$ is diagonal, with $\mu_{xx}$ and $\mu_{yy}$ describing glide and climb mobilities, respectively. We expect that the mobility tensor itself will have a turgor pressure dependence, as the resulting forces can lower the activation barriers of the various biochemical pathways involved in the process.
Thus, the observed velocity of the strand extension machinery should depend on the excess pressure for rod-shaped bacteria.
%
%In experiments, the velocity of the labeled glycan-strand-extending  proteins does not depend strongly on the growth medium (but does show a strong dependence on the temperature) \cite{shaevitz}. This observation suggests that, as assumed here, the dislocation dynamics is governed by energetic considerations, and as such should be quite robust to the biological details such as the growth medium (which can, however, affect the total number of active dislocations, and hence the overall growth rate of the bacteria).

\subsection {The role of interactions}
\label {interaction_sect}
  In condensed matter physics, the long-range elastic interactions between dislocations can have important consequences, and have recently been suggested to lead to glassy effects and non-thermal, heavy-tailed, dislocation velocity distributions \cite {ispanovity,ispanovity2,goldenfeld}.

  To illuminate their importance in the biological context, we have solved for the interaction energy of two dislocations with antiparallel Burgers vectors $\pm \vec{b}$ on an infinite cylinder, separated by a distance $x$ along the long-axis and $y$ along the circumference \cite{supplement_dislocations}.

\begin{eqnarray}
&  & E(x,y)= \frac{Ab^{2}h}{2} [i\frac{x}{2R}\rm {csc}(\frac{y}{2R})\rm {sinh}(\frac{x}{2R})\rm {csch}[\frac{x-iy}{2R}] \nonumber \\
 &  & +\rm {log}(\frac{2R}{b}\rm {sinh}[\frac{x-iy}{2R}]] +C.C. \label{energy}
\end{eqnarray}
Here, $R$ the radius of the cylinder, $h$ the cell wall thickness, and $A\equiv\frac{\mu}{2\pi(1-\nu)}$, with $\mu$ the shear modulus and $\nu$ the Poisson ratio (we assume isotropic elasticity theory, for simplicity). For distances $x \gg R$ along the axis of symmetry, Eq. (\ref{energy}) shows that the interactions fall off exponentially \cite{bruinsma2}.
For distances much smaller than the cylinder radius, the interaction energy reduces to its form in two-dimensions, as expected. Clearly, the relevant scale for the interaction energy is $Ab^2 h$.
Fig. 1a in the Supplementary Materials shows an example of the equal energy contours. Close to the origin, taking a cut parallel to the $x$
axis would give a graph with two minima, corresponding to the two
dislocations sitting at a 45 degree angle - this is
expected, since the physics close to the origin should be insensitive to the finite
cylinder radius: this arrangement is indeed the stable configuration
of two dislocations in Euclidean space, when climb processes are prohibited
\cite{bruinsma}.
Fig. 1b  in the Supplementary Materials illustrates how interactions affect an activated dislocation (\emph{i.e.}, one with strand-extending machinery attached) attempting to move upwards from a fixed, unactivated defect.

Let $G$ represent the combined effect of the chemical forces provided by enzyme-mediated polymerization and those caused by turgor pressure, as described by Eq. (\ref{fy}). The sign of this force depends on the sign of the Burgers vector, and acts in the direction which corresponds to elongation of the existing strands. For a vertical separation $y$ in the range $b<y\ll 2\pi R$, we can use Eq. (\ref{energy}) and add a term $E_{field}=-Gy$ corresponding to the force $G$. This leads to the potential energy $U(y)=2E_c+(Ab^2h)\log(y/b)-Gy$, where $2E_c$ is the energy at separation $y=b$ and $G=0$. The logarithmic interaction term clearly leads to a long range elastic force that falls off slowly, like $1/y$.   Upward motion will require thermal activation unless $y>y^*=Ab^2h/G$. Escape from this logarithmic binding will be relatively easy for $G$ large enough to insure $y^* \leq b$, \emph{i.e.}, $G>G^* \equiv Abh$  . Healthy, growing cells in rich nutrient environments are likely to be in this regime and thus escape the confining effect of elastic forces. Elastic interactions will nevertheless become important when $G$ is diminished by osmotic shock, nutrient deprivation, or by application of antibiotics that interfere with the powerful strand extension machinery.

Fig. 2 illustrates the dynamics of several tens of active dislocations under the influence of a circumferential ``field" or driving force $G$ on each dislocation with pairwise elastic interactions given by Eq. (\ref{energy}), moving in the disordered background created by inactive dislocations, to be elaborated upon below. For a weakly disordered system, one obtains ordered circumferential motion (see Fig. 3 in the Supplementary Material), while for a disorder level chosen according to the biological parameters, the elastic interactions are strong enough to occasionally pin the motion.
This could be related to pauses in the circumferential motion observed in recent experiments \cite{garner}. The condition for pinning to be important follows from comparing the typical elastic interaction force with the nearest dislocation a distance $r_{nn}$ away (which is highly likely to be inactive) and the net driving force $G$. Pinning is expected to occur when:
\be \frac{Ahb^2}{r_{nn}} =Ahb^2 n^{1/2} > G, \ee where $n$ is the dislocation areal density. With parameter estimates for \emph{E. coli} (see the section ``Estimation of the biological parameters'" in the Supplementary Material) we find that the ratio $\chi \equiv Ahb^2 n^{1/2}/G$ is of the order of unity, suggesting that pinning events are likely to occur. However, due to the finite processivity of the active dislocations, pinning can be bypassed by the growth dynamics, as the elongation machinery can drop off a pinned dislocation and continue the strand elongation elsewhere. However for biological conditions (nutrient depletion or low turgor pressure) such that $\chi \gg 1$, pinning events could become much more important.

\begin{figure}[b!]
\includegraphics[width=0.5\textwidth]{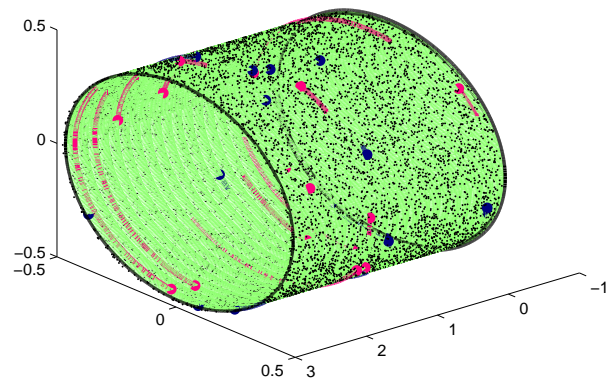}
\caption { Numerical simulation of the dynamics of 30 active dislocations, with simulation parameters matching the biological ones.  The dislocations are driven by a combination of chemical forces and turgor pressure in the circumferential direction, proportional to the sign of the Burgers vector. 10,000 inactive dislocations create a disordered potential for the motion of the active dislocations. The temperature is zero and
$\frac{2\pi R G}{Ab^2h}=450$. The $x$ and $y$ coordinates of the dislocations were chosen randomly and uniformly along the axes, and $W/L=\frac{2 \pi R}{L}=1$. The climb to glide mobility ratio is $\mu_c/\mu_g=10$. The red and blue points mark the starting positions of the $\pm b$ dislocations, and the red and blue lines correspond to their trajectories. The black circles mark the end of the cylinder. A snapshot of the simulation is taken after $t=0.7 \frac{W}{\mu_c G}$. The numerous inactive dislocation (marked as black dots) create a disordered energy landscape which can trap the active dislocations and inhibit further growth. \emph{In the Supplementary Material, further details regarding the simulation are given.}} \label {dynamics_fig1}
\end{figure}

%
%For a weak field, the dynamics is complex due to long ranged interactions between the dislocations, while for strong fields one obtains orderly motion around the circumference.
Evidently, experimental measurements of interaction-induced correlations between moving dislocations could provide a useful probe of the elastic properties of the peptidoglcan.
Estimates of the relative strength of the interactions compared to the other forces involved in the problem are provided in the Supplementary Material, and discussed below.

%The elastic constants of the peptidoglycan mesh are believed to be of the order of $2\cdot 10^7 Pa$ \cite{yao}, which leads to a typical scale of $??? erg$ associated with the core energies of the dislocations.
\section{Dislocation dynamics}

\begin{figure}[b!]
\includegraphics[width=0.5\textwidth]{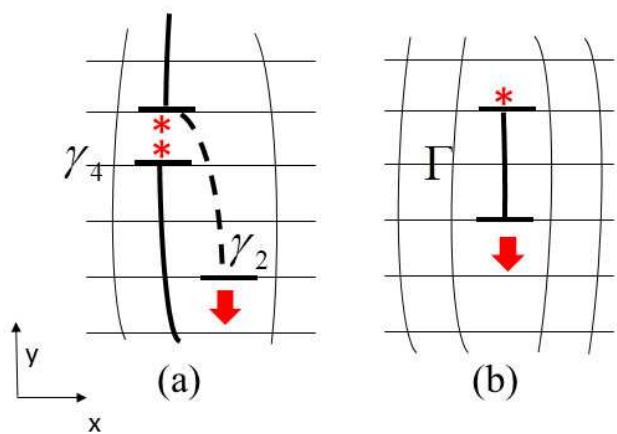}
\caption {(a) Active dislocation pulled past an inactive one by the enzymatic machinery. In this case an initial (inactive) dislocation pair could be created by an enzyme that cuts glycan bonds (rate constant $\gamma_4$). The elongation machinery then assembles around an inactive dislocation turning it into an active one (rate constant $\gamma_2$). (b) Direct insertion of a glycan strand fragment to create an active and inactive dislocation pair (rate constant $\Gamma$).} \label {growth_modes}
\end{figure}

Eq. (\ref{growth}) expresses the rate of change of the bacterial surface area $S(t)$. To proceed further, we must account for the dynamics of the number of active and inactive dislocations, controlled by a variety of interesting biological processes:

1. Dislocation pair creation: it is plausible that the nucleation of new growth sites should be proportional to the cylinder's area (see Fig. 3b). For growth under rich nutrient conditions, we expect the area \emph{density} of peptidoglycan growth sites (\emph{i.e.}, activated dislocations) to be constant, which requires a source of new dislocations as the bacteria elongates with fixed radius.
Dislocation pairs could in principle be created by thermal activation \cite{amir_nelson_PRE,bruinsma}, although this process is likely to be rare in a covalently bonded peptidoglycan mesh.
Alternatively, a dislocation pair can form by breaking bonds in the peptidoglycan mesh, via an enzymatic process or via external damage by radiation or severe stresses \cite{bound_dislocation_comment}.

2. The finite processivity of elongation enzymes:  the elongation machinery can fall off an active dislocation, and render it inactive. In an elegant recent work \cite{jacobs}, finite processivity was implicated in straightening of an initially curved \emph{Caulobacter crescentus}, where the working hypothesis was that the growth occurs via partial ``hoops" extending perpendicular to the long axis of the rod shaped bacteria, as described in Refs. \cite{shaevitz,garner,escobar}. For \emph{C. crescentus}, processivity was such that on average 1/5 of the circumference is traversed before the machinery falls off \cite{jacobs}. It would be interesting to perform a similar study for \emph{E. coli} or \emph{Bacillus subtilis}. Growing filamentous \emph{E. coli} in curved microchambers was also demonstrated \cite{whitesides}, proving the feasibility of such an experiment for \emph{E. coli}.

3. The activation of an inactive dislocation: the reverse process assembles the elongation machinery onto an existing (static) glycan strand end and turns it into an actively growing dislocation. The rate of this process is proportional to the number of inactive dislocations.

4. Annihilation processes: two dislocations, both active or one inactive, can ``collide", completing a full circumferential strand. The elongation machinery could then immediately continue to move and grow the cell wall (as in Fig. 3a), in which case the annihilation process plays no role. Alternatively, it could tie the strands together, dislodge and become available for a new inactive dislocation. In complete analogy to the theory of chemical reactions, the rate of this latter process per unit area is proportional to the product of the density of the active dislocation and that of the total number of dislocations.

Upon collecting together these four processes, we describe the dynamics of the numbers of active and inactive dislocations on a cylindrical wall of surface area $S(t)$ by supplementing Eq. (\ref{growth}) with the following:

\begin{equation}
\frac{dN_{ac}}{dt}=\Gamma S-\gamma_1N_{ac}+\gamma_2N_{in}-\gamma_3 v N_{ac}(n_{ac}+n_{in}), \label{growth1}
\end{equation}

\begin{equation}
\frac{dN_{in}}{dt}=\Gamma S+\gamma_1 N_{ac}-\gamma_2 N_{in}+2\gamma_4 S -\gamma_3 v n_{ac}N_{in}. \label{growth2}
\end{equation}
In the above, $N_{ac}$ and $N_{in}$ denote the number of active
and inactive dislocations, and $n_{ac}=N_{ac}/S$ and $n_{in}=N_{in}/S$ denote their densities. The term $\Gamma$ arises from the nucleation of dislocation pairs, one active and one inactive, by a process like that shown in Fig. 3b, while $\gamma_1$ represents the processivity and $\gamma_2$ describes the activation of an inactive dislocation (see Fig. 3a). The rate $\gamma_3$ represents the annihilation of a dislocation pair (which is typically proportional to the velocity \cite{amir_nelson_PRE}). Finally, $\gamma_4$ describes creation of tightly bound inactive dislocation pairs by an enzyme which adds to the reservoir of activatable dislocations by cutting covalent bonds (Fig. 3a). Note that no spatial structure is present in the above equations: the dynamics is ``mean-field" like, \emph{i.e.}, it assumes that the dislocations are well-mixed and that their density is approximately uniform on the cylinder \cite{nelson_review_comment}. A more sophisticated theory would explore the effects of spatial heterogeneities on the dynamics \cite{amir_nelson_PRE}.

We search for steady state densities $n_{ac}$ and $n_{in}$ with these equations, taking into account Eq. (\ref{growth}). The
steady state dislocation densities then obey:

\begin{equation}
n_{ac}^{2}vb=\Gamma-\gamma_1 n_{ac} + \gamma_2 n_{in} -\gamma_3 v n_{ac}(n_{ac}+n_{in}),
\end{equation}

\begin{equation}
n_{in}n_{ac}vb={\Gamma}+\gamma_1 n_{ac}-\gamma_2 n_{in}+2\gamma_4-\gamma_3 v n_{ac}n_{in}.
\end{equation}

The terms on the left arises because the dislocations are automatically diluted as the cell wall grows.
From Eq. (\ref{growth}), we see that the length of the bacteria $L(t)=\frac{S(t)}{2\pi R}$ grows exponentially in the steady state,

\begin{equation}
L(t)=L_{0}e^{n_{ac}vbt}.
\end{equation}

While we cannot rule out the possibility of cell cycle effects in the elongation rate, our simplified model is consistent with the existing data for exponential growth, for bacteria growing in one dimensional channels with abundant nutrients \cite{jun}. As discussed below, the $G$ force driving dislocation climb enters our dynamical equations, through (1) $\Gamma(G)$, the escape rate of bound dislocations, (2) the velocity $v=v(G)$, (3) through the capture cross-section embodied in $\gamma_3=\gamma_3(G)$, and (4) the creation rate of new dislocations. In particular $\Gamma$ and $\gamma_4$, might depend on $G$,  as it could change the barriers for thermal activation (see \cite{wagner_plasticity} for a related discussion).

%This could come about as a result of two possible scenarios:

%1. $ \gamma_4 \gg \Gamma -  \gamma_1 n_{ac} +  \gamma_2 n_{in} - \gamma_3 n_{ac}(n_{ac}+n_{in})$.

%2. $\Gamma \approx   \gamma_1 n_{ac} +  \gamma_2 n_{in}- \gamma_3 n_{ac}(n_{ac}+n_{in})$, even for $ \gamma_4=0$.

%The latter scenario, however, demands fine tuning, and does not seem to be plausible. We therefore conclude that $c$ is large and governs the creation of new inactive dislocations as the bacteria elongates. Under this assumption, the number of inactive dislocations is given by:
%\be n_{in} \approx .\ee

%Thus, the balance of these two processes will yield a similar equation
%to the one we had before, but with $\lambda_{recapture}\propto N_{active}n_{total}/D$.

%These results imply that at high densities of dislocations, where $\tau$ (which is inversely proportional to the density) would be
%smaller than the reciprocal escape rate, the growth rate will be independent
%of the density! This occurs since the "mean free path" becomes
%shorter as the density increases. In the opposite regime, of a low
%density of dislocations, the growth would be proportional to the density,
%as intuitively expected.

%In fact, if the creation of new elongation machinery is eliminated, then ???

\section{The biological parameters}

In the Supplementary Material, we estimate the relevant biological parameters for the case of \emph{E. coli}, in order to elucidate which of the previous physical processes should be important. We estimate the energy of creating a new dislocation pair using the measured elastic constants, and find that it is much larger than the temperature. Using the measured turgor pressure, we find that the force exerted on each of the dislocation by the stresses arising from it is similar in strength to those due to the biochemical reaction involved in the elongation process. This suggests that changing the turgor pressure or alternatively, putting the bacteria under mechanical stresses, can result in significant changes in the velocity of the dislocations. The density of \emph{inactive} dislocations can be estimated from the strand length distribution, and is found to be much larger than that of \emph{active} dislocations. Finally, the typical elastic interaction forces between nearby \emph{active} dislocations are negligible compared to the driving force in the circumferential direction, but the elastic interactions between the \emph{active} dislocations and the numerous \emph{inactive} ones are comparable to the driving force, suggesting that the possibility of dislocation pinning, discussed earlier, could occur.

\section{Bacterial elongation rate}

The assumption $n_{in} \gg n_{ac}$ simplifies Eqs. \ref{growth1} and \ref{growth2}. We explore here two possible scenarios:

 1. \emph{A two-stage process:} An enzyme cuts glycan links in the mesh to produce a pair of inactive dislocations, while other proteins recruited by the MreB then activate a small subset of them, accounting for the small subset of active dislocations (see Fig. 3a). Thus far, the annihilation of dislocations does not seem to be observed in experiments \cite{garner1}, suggesting that we could also ignore $\gamma_3$. With these assumptions, we can simplify Eqs. (\ref{growth1}) and (\ref{growth2}) in the steady state: $ n_{ac}^{2}vb=\gamma_2 n_{in}- \gamma_1 n_{ac} \label {scenario1} $,
$ n_{in}n_{ac}vb=\gamma_1 n_{ac}-\gamma_2 n_{in}+2\gamma_4.$

Upon adding these equations and using $n_{in} \gg n_{ac}$ we obtain:
$n_{in}n_{ac}vb=2\gamma_4$ ,which when combined with the previous equations leads to a cubic equation for $n_{ac}$, the density of active dislocations:

\be n^3_{ac} b^2 v^2 + \gamma_1 n^2_{ac} bv =2 \gamma_2 \gamma_4. \label{scaling1} \ee

For $b v \gamma_2 \gamma_4 \gg \gamma^3_1 $ (short processivity of the elongation machinery) we find $ n_{ac}= (\frac{2 \gamma_2 \gamma_4}{b^2 v^2 } )^{\frac{1}{3}}$, while for $ b v \gamma_2 \gamma_4 \ll \gamma^3_1 $ we get $n_{ac}= \sqrt{\frac{2 \gamma_2 \gamma_4}{b v \gamma_1}} \label{scaling2}$.

It is useful to rewrite these scaling relations in terms of the exponential growth rate $\lambda = n_{ac} vb$, where $L(t)=L_0 e^{\lambda t}$. This leads to the relations:
$ \lambda = ({2 b v \gamma_2 \gamma_4})^{\frac{1}{3}}$ , for short processivity, and:
 $\lambda= \sqrt{\frac{2 bv\gamma_2 \gamma_4}{ \gamma_1}}$  in the opposite regime.  An experimentally accessible control (\emph{e.g.}: temperature) for which one of the microscopic processes (e.g.: $\gamma_1$) is much more sensitive than the others, would provide a way to test these scaling relations without experimentally measuring \emph{all} of the microscopic rates.

2. \emph{Direct insertion of a nascent strand:} Another possible scenario is that dislocation pairs are formed by inserting material into the mesh (but on new sites, between the existing mesh, see Fig. 3b), which are further acted upon by the elongation machinery to produce the growth. If this is the dominant mechanism we can neglect $\gamma_4$. If in addition we neglect the annihilation term, as discussed previously, we obtain the steady state equations: $ n_{ac}^{2}vb=\Gamma-\gamma_1 n_{ac} + \gamma_2 n_{in}$, $n_{in}n_{ac}vb={\Gamma}+\gamma_1 n_{ac}-\gamma_2 n_{in}$.

Upon again invoking $\frac{n_{in}}{n_{ac}}\gg1$, we find that $n_{ac}n_{in}bv \approx 2 \Gamma$, which leads to a quadratic equation for $n_{ac}$:

\be \gamma_1 n_{ac}^2 - \Gamma n_{ac} -\frac{2\gamma_2 \Gamma}{bv}=0. \label{scaling3} \ee
For $\gamma_1 \gg \frac{\Gamma bv}{ \gamma_2}$, we obtain $n_{ac} \approx \sqrt{\frac{2\gamma_2 \Gamma}{\gamma_1 bv}}$.

%Figure \ref{dynamics_fig1} illustrates the dynamical processes described above, and the interplay of active and inactive dislocations.

Another useful limit can be obtained by setting $\gamma_2=0$. In this limit active dislocations can only be created by the direct insertion, and inactive dislocations to not play a role. In this case, which was studied in Ref. \cite{nelson_review}, the two equations decouple, and one obtains a quadratic equation for $n_{ac}$.

We anticipate that experiments will soon determine the density of active dislocations, thus allowing tests of the above scaling relations. As explained previously, parameters such as the processivity could be experimentally determined independently, allowing additional checks. Another interesting biological scenario is one where the creation of new dislocations is abruptly stopped, by disabling the relevant protein. In this case $\gamma_4=\Gamma=0$, and subsequent steady-state growth is \emph{linear} rather than exponential, since the total number of dislocations is constant in this case. From Eqs. (\ref{growth1}) and (\ref{growth2}) we can determine the typical transient time $\tau$ for the growth to cross over to the linear regime is, $\tau = \frac{1}{\gamma_1+\gamma_2}$.
This and related experiments could provide further insight into the microscopic rate constants discussed above.

%
%
%
%we should be able to see the effects of interaction on the dislocation dynamics.  However, if this is the case it then it seems that the Turgor pressure is not sufficiently large to overcome the energy barrier associated with the creation of two dislocations. This might imply that we have indeed $\Gamma=0$ in Eq. (\ref{growth1}), such that the elongation machinery does not produce pairs of dislocations and only operates on previously existing static dislocations, \emph{i.e.}, places in the mesh where another protein cut the peptide links to create an open strand.
%
%%\emph{Processivity of an active dislocation .-}  The processivity of the elongation machinery is ????, which implies that ???.

\emph{Protein regulation and exponential growth.-}
Cell wall growth by itself is unlikely to be a bottleneck process that limits bacterial growth \cite{hagen}. In fact, bacterial cells probably regulate the concentration of all proteins, DNAs, RNAs \emph{etc.} involved in the cell growth, to yield a coordinated set of doubling times determined by richness of the media, temperature \emph{etc}. In our model, parameters such as the dislocation velocity, driving force $G$ and density of active dislocations will be determined by the regulatory network that controls cell growth. One important feedback mechanism is surely the ``geometrical dilution": as the cell wall grows, the concentrations of all the proteins in the cells are diminished. If we assume ribosome production is the limiting factor of the growth \cite{hwa} (ribosomes constitute of most of the dry weight of the cell), and the relevant proteins are maintained at constant \emph{concentration}, the cell wall dynamics discussed here would be consistent with the ribosomal constraints. Regulation of concentration is appealing, as it is a local and robust mechanism, consistent with the biological ideas about self-organization \cite{camazine}. Locality is also connected with the appearance of the surface area in Eqs. (\ref{growth1}) and (\ref{growth2}), as discussed earlier. Even if some more complicated biological process is involved, we expect predictions such as Eqs. (\ref{scaling3}) to hold; one should be able to check them directly by measuring the rates of the underlying individual processes, and by engineering sudden changes in rates such as $\gamma_1$, $\gamma_2$ or $\gamma_4$. Finally, we note that the aforementioned geometric dilution is quite generic, and will appear in every case where the cell volume grows. In Ref. \cite{alon}, section 2.4, this dilution is accounted for by adding a term to the protein degradation. In our case, however, the dilution of the \emph{active} dislocations is not proportional to their concentration (as is usually the case with protein degradation), but enters \emph{quadratically}. This comes about since the active dislocations themselves are responsible for the change in surface area.

\section{Conclusions and outlook}

In this paper we used the continuum elastic theory to account for recent experiments on the growth of single bacteria. Our theoretical framework exploits the partially ordered structure peptidoglycan mesh to obtain an economical description of the growth dynamics in terms of dislocation defects. The forces acting on these defects lead to predominantly \emph{climb} dynamics, quite different from those typically found in condensed matter physics and materials science. Our results suggest that osmotic pressure could affect the measured dislocation velocities, and that the elastic interactions could lead to pinning of the dislocation motion. Consideration of various processes associated with dislocation motion, and their creation and annihilation, leads to a coupled set of ordinary differential equations that govern the dynamics of the number of active and inactive dislocations, and the overall elongation rate. Solving these under conditions of abundant nutrients leads to exponential growth, consistent with the experimental observations. Transients and other modes of growth are also discussed. In the future, it would be interesting to compare and contrast bacterial elongation with pollen tube growth, another case where the growth occurs on a cylinder with an approximately constant radius \cite{maha}. One might also speculate that planar arrangements of cells in, say, embryos, grow by cell divisions sited preferentially in dislocation cores. The resulting dislocation climb takes place in array of living cells.  Another natural extension of this approach might be to model the growth of approximately spherical cells \cite{nelson_review}, such mutant \emph{E. coli}. Finally, it would be useful to extend this theory to the thicker cell walls of gram-positive bacteria, which experimentally show similar phenomenology \cite{garner, escobar}.

%% == end of paper:

%% Optional Materials and Methods Section
%% The Materials and Methods section header will be added automatically.

%% Enter any subheads and the Materials and Methods text below.
%\begin{materials}
% Materials text
%\end{materials}

%% Optional Appendix or Appendices
%% \appendix Appendix text...
%% or, for appendix with title, use square brackets:
%% \appendix[Appendix Title]

\begin{acknowledgments}
It is our pleasure to thank E. Berg, E. Garner, B. I. Halperin, J. H. Hutchinson, G. J. Jensen, F. Spaepen and S. Wang for useful discussions. We are grateful to P. Hohenberg, T. Hwa and B. Shraiman for a critical reading of the manuscript.
This work was supported by the National Science Foundation via Grant DMR1005289 and through the Harvard Materials Research Science and Engineering Laboratory, through Grant DMR0820484. A.A. was supported by a Junior Fellowship of the Harvard Society of Fellows. \end{acknowledgments}

%% PNAS does not support submission of supporting .tex files such as BibTeX.
%% Instead all references must be included in the article .tex document.
%% If you currently use BibTeX, your bibliography is formed because the
%% command \verb+\bibliography{}+ brings the <filename>.bbl file into your
%% .tex document. To conform to PNAS requirements, copy the reference listings
%% from your .bbl file and add them to the article .tex file, using the
%% bibliography environment described above.

%%  Contact pnas@nas.edu if you need assistance with your
%%  bibliography.

% Sample bibliography item in PNAS format:
%% \bibitem{in-text reference} comma-separated author names up to 5,
%% for more than 5 authors use first author last name et al. (year published)
%% article title  {\it Journal Name} volume #: start page-end page.
%% ie,
% \bibitem{Neuhaus} Neuhaus J-M, Sitcher L, Meins F, Jr, Boller T (1991)
% A short C-terminal sequence is necessary and sufficient for the
% targeting of chitinases to the plant vacuole.
% {\it Proc Natl Acad Sci USA} 88:10362-10366.

%% Enter the largest bibliography number in the facing curly brackets
%% following \begin{thebibliography}

%\bibliography{bacteria}

\begin{thebibliography}{10}
\bibitem{john_comment}
\newblock The criterion for the internal pressure at which this effect can be neglected will be quantitatively discussed in future work.
\end{thebibliography}
%\bibliographystyle{pnas2009}

%
%

\setcounter{equation}{0}
\setcounter{figure}{0}

\newpage
\huge{Supplementary Material}
\vspace{1 cm}

\large
\textbf{Dislocation interactions on a cylinder}
\normalsize

Consider edge dislocations on the surface of a cylinder, with radius $R$ and infinite length. Let us denote the coordinate along the symmetry axis by $x$, and the other by $y$, \emph{i.e.}, $y$ is periodic with a period of $W \equiv 2\pi R$. We shall assume throughout that the Burgers vector of all dislocations point in the $\pm x$ direction, with magnitude $b$.

 We will not consider out-of-plane buckling, which is suppressed by the high turgor pressure inside the bacteria \cite{john_comment}. Confined to the surface of the cylinder, the Laplacian operator is equivalent to that in an infinite two-dimensional space together with the periodicity
requirement. Therefore, to find the stress exerted by a dislocation at position $(x_{0},y_{0})$ on another at position $(x,y)$, we have to solve the bi-harmonic equation for the Airy stress function \cite{hirth} on a strip with periodic boundary conditions.

The stresses without the periodicity requirement are given by (up to a sign depending on the direction of the Burgers vector):

\begin{equation}
\sigma_{xx}=-Abh(y-y_{0})[3(x-x_{0})^{2}+(y-y_{0})^{2}]/r^{4},\label{sigma_xx}
\end{equation}

\begin{equation}
\sigma_{yy}=Abh(y-y_0)[(x-x_0)^{2}-(y-y_0)^{2}]/r^{4},
\end{equation}

 and
\begin{equation}
\sigma_{xy}=Abh(x-x_{0})[(x-x_{0})^{2}-(y-y_{0})^{2}]/r^{4},\label{sigma_xy}
\end{equation}
 with $r^{2}=(x-x_{0})^{2}+(y-y_{0})^{2}$ and $A\equiv\frac{\mu}{2\pi(1-\sigma)}$,
with $\mu$ the shear modulus and $\sigma$ the Poisson ratio. Note that these stresses are in two-dimensions, and hence their units are force per unit length.

To describe stresses on a cylinder, the dislocation at $(x_{0},y_{0})$ must be duplicated at intervals of $W=2\pi R$ in the $y$ direction. We choose, without loss of generality, $x_{0}=y_{0}=0$. Therefore the stresses it generates at $(x,y)$ are given by:
\begin{equation}
\sigma_{xx}^{cylinder}=\sum_{k=-\infty}^{\infty}\frac{-Abh[y-kW][3x^{2}+(y-kW)^{2}]}{[x^{2}+(y-kW)^{2}]^{2}},
\label{sum1}
\end{equation}

\begin{equation}
 \sigma_{yy}^{cylinder}= \sum_{k=-\infty}^{\infty}\frac{Abhy[x^{2}-(y-kW)^{2}]}{[x^{2}+(y-kW)^{2}]^{2}},
\label{sum2}
\end{equation}

\begin{equation}
 \sigma_{xy}^{cylinder}= \sum_{k=-\infty}^{\infty}\frac{Abhx[x^{2}-(y-kW)^{2}]}{[x^{2}+(y-kW)^{2}]^{2}}.
\label{sum3}
\end{equation}

The force on an additional dislocation at $(x,y)$ due to this stress will then be given by the Peach-Koehler force \cite{peach,hirth}:

\begin{equation}
F_{x}=b\sigma_{xy}^{cylinder},\label{force_x}
\end{equation}

\begin{equation}
F_{y}=-b\sigma_{xx}^{cylinder}.\label{force_y}
\end{equation}

Although the stress component $\sigma_{yy}$ does not enter the forces in this geometry, we will also evaluate it, for completeness.
%
%\vspace{10 mm}
%
%\large
%\textbf{Summing the stresses}
%\normalsize
%
%\label{forces}

The sums of Eqs. (\ref{sum1})   -(\ref{sum3}) can be evaluated using the Sommerfeld-Watson transformation \cite{walker}. We illustrate this method for the first sum. Since the function $g(z)={\rm {cot}(\pi z)}$ has only simple poles of residue unity that lie on the $x$ axis at integer values, the sum in Eq. (\ref{sigma_xx}) can be written as the complex contour integral:

\begin{equation}
\oint_{C}g(z)f(z)dz,
\end{equation}
 with:

\begin{equation}
f(z)=-Abh\frac{(y/W+z)[3(x/W)^{2}+(y/W+z)^{2}]}{W[(x/W)^{2}+(y/W+z)^{2}]^{2}}.
\end{equation}
 Since $f(z)\sim1/z$ at large distances from the origin, we can deform the contour so that it captures only the poles of $f(z)$ (the integral on the circle at infinity vanishes even though the decay is only $\sim1/z$, due to the ${\rm {cot}(\pi z)}$ term).

The function $f(z)$ can be rewritten as:

\begin{equation}
f(z)=-(Abh/W)\frac{(y/W+z)[3(x/W)^{2}+(y/W+z)^{2}]}{(z+(y+ix)/W)^{2}(z+(y-ix)/W)^{2}}.
\end{equation}
 Clearly, it has two poles, each of order 2. Upon summing the residues we find:

\begin{eqnarray}
& &\sigma_{xx}=\oint_{C}g(z)f(z)dz= \nonumber \\
& &\frac{iAbh\pi^{2}x}{2W^{2}}{\rm {csc}^{2}(\pi(y-ix)/W)} \nonumber \\
& & -\frac{\pi Abh}{2W}{\rm {cot}(\pi(y+ix)/W)} +C.C. \label{sigma_xx_cylinder}
\end{eqnarray}

 In a similar fashion one obtains:

\begin{eqnarray}
& & \sigma_{yy}=-\frac{iAbh\pi^{2}x}{2W^{2}}{\rm {csc}^{2}(\pi(y-ix)/W)} \nonumber \\
& & -\frac{\pi Abh}{2W}{\rm {cot}(\pi(y+ix)/W)}+C.C. , 
\end{eqnarray}

and

\begin{equation}
\sigma_{xy}=-\frac{Abh\pi^{2}x}{2W^{2}}{\rm {csc}^{2}(\pi(y-ix)/W)}+C.C.\label{sigma_xy_cylinder}
\end{equation}
 %stresses in Eqs (\ref{sigma_xx}) and (\ref{sigma_xy})

%The forces are plotted in Figs. (\ref{Fx}) and (\ref{Fy}).
Upon integration, one obtains Eq. (5) in the main text, describing the interaction energy of two dislocations on a cylinder as a function of their separation. The energy landscape, with and without an additional constant force along the circumference, is shown in Fig. 1.

%
%~
%\begin{figure*}[b!]
% \includegraphics[width=0.5\textwidth]{Fx.eps} \caption{The component of the force along the cylinder axis $x$, as a function
%of $x$ and $y$.}
%\label{Fx}
%\end{figure*}
%
%
%~
%\begin{figure*}[b!]
% \includegraphics[width=0.5\textwidth]{Fy.eps} \caption{The component of the force perpendicular to the cylinder axis $x$,
%as a function of $x$ and $y$.}
%\label{Fy}
%\end{figure*}
%

\begin{figure}[b!]
\includegraphics[width=0.45\textwidth]{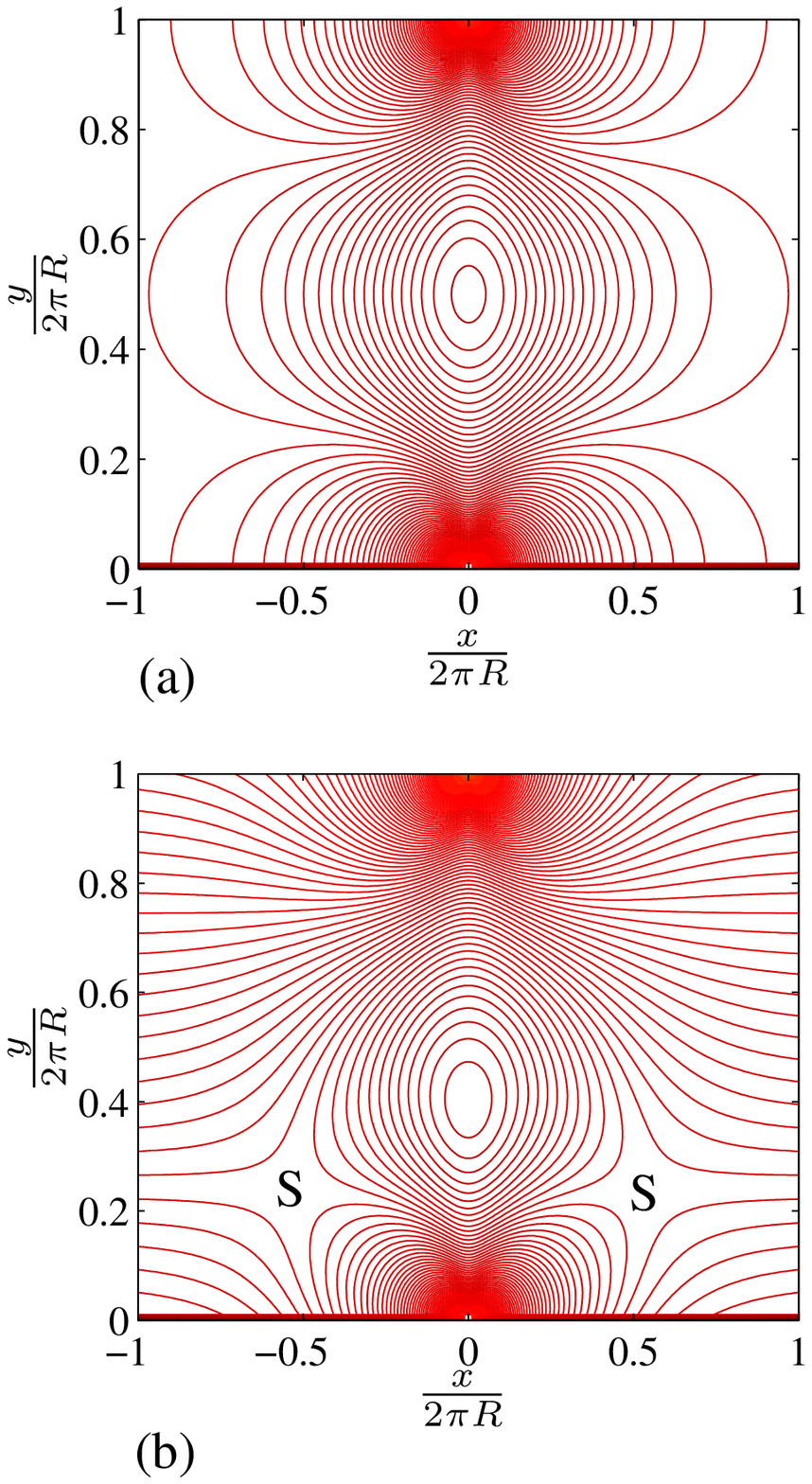}
\caption{(a) Contours of $E(x,y)$ for the interaction energy of two dislocations on an infinite cylinder, as given by\ Eq. (5) in the main text. The $y$ axis is along the circumference, and the $x$ axis is along the axis of symmetry. (b) Energy contours of $E(x,y)-Gy$ with $\frac{2\pi R G}{Ab^2h}=1$. As explained in the main text, a combination of chemical forces and turgor pressure would create such a force $G$ on an edge dislocation with a Burgers vector along $x$. This force creates two saddle points denoted by $S$ in the potential energy landscape. If thermal activation were the dominant nucleation process, the properties of these saddle points would control creation of dislocation pairs \cite{amir_nelson_PRE}. } \label{energy_landscape}
\end{figure}

%
%\begin{figure}
%\centering
%\mbox{(a) \subfigure{\includegraphics[width=0.25 \textwidth]{G_0.eps}}\quad
%\subfigure{ (b) \includegraphics[width=0.25 \textwidth]{G_1.eps} }}
%\caption{(a) Contours of $E(x,y)$ for the interaction energy of two dislocations on an infinite cylinder, as given by\ Eq. (5) in the main text. The $y$ axis is along the circumference, and the $x$ axis is along the axis of symmetry. (b) Energy contours of $E(x,y)-Gy$ with $\frac{2\pi R G}{Ab^2h}=1$. As explained in the main text, a combination of chemical forces and turgor pressure would create such a force $G$ on an edge dislocation with a Burgers vector along $x$. This force creates two saddle points denoted by $S$ in the potential energy landscape. If thermal activation were the dominant nucleation process, the properties of these saddle points would control creation of dislocation pairs \cite{amir_nelson_PRE}. } \label{energy_landscape}
%\end{figure}

\vspace{10 mm}

\large
\textbf{Numerical simulation of a system of interacting dislocations}
\normalsize

To study the dynamics of many interacting dislocations, we numerically simulate a system of 30 active, overdamped edge dislocations moving on the surface of a cylinder under the influence of a disordered potential created by 10,000 inactive dislocations. The parameters are chosen according to the estimates of the biological parameters, described in the main text. Their Burgers vector $\pm b$ is along the $x$ axis, which coincides with the cylinder's long axis, and we insert equal numbers of dislocations with positive and negative $b$. The initial positions are random along the circumference, and uniformly distributed along the cylinder on the interval $[0,d]$, with $d \approx 1500 b$, for both active and inactive dislocations. The force $G$ on a dislocation consists of two terms: one arises from the $G_{pressure} \equiv b\sigma_{xx}$ term due to the turgor pressure inside the bacteria, the second due to the energy change $u$ associated with the biochemical reactions involved in the elongation process, as discussed in the main text. Another contribution to the force arises from the elastic interactions between the dislocations, which (for an infinitely long cylinder) are given by Eqs. (\ref{force_x}),(\ref{force_y}),(\ref{sigma_xx_cylinder}) and (\ref{sigma_xy_cylinder}). The number of inactive dislocations, $N_{in}$, is 10,000 for the most biologically relevant case, shown in Fig. 3 of the main text. A cutoff $F_c$ on the maximal elastic forces exerted by a dislocation is imposed, associated with the short distance cutoff, which we take to be of order $b$, the lattice spacing. All simulations are run at zero temperature, and solved by direct integration of the equations of motion. The total simulation running time is denoted by $T_s$. An important parameter choice in the simulations involves the mobility tensor $\mu_{ij}$ defined by Eq. (4) of the main text. We take this tensor to be  diagonal in the $(x,y)$ coordinate system defined in Fig. 1 of the main text, with a glide mobility $\mu_g = \mu_{xx}$ and climb mobility $\mu_c=\mu_{yy}$. We approximate the predominantly climb motion set up by the elongation enzymes via the choice $\mu_c = 10 \mu_g$. The small non-zero glide mobility allows us to model the fluctuations of the motion along the cylinder's long axis, which could arise from disorder in the peptidoglycan mesh.

To illustrate further the effects of elastic interactions between dislocations, we eliminated all inactive dislocations and simulated the dynamics of 30 active dislocations \emph{strongly interacting} on a cylinder, again with a Burgers vector $\vec{b}=\pm x$ . The results are shown in Figs. 2 and 3, for two values of the driving force $G$, differing by a factor of 10. For the stronger driving field the elastic interactions are less important, and one obtains ordered circumferential motion. For the weaker field, the attractions between dislocation of opposite sign of the Burgers vector are strong enough such that they occasionally overcome the driving force and meet, leading to their mutual annihilation. A detailed study of the annihilation process of two interacting dislocations will be given elsewhere \cite{amir_nelson_PRE}.

\begin{figure}[b!]
\includegraphics[width=0.5\textwidth]{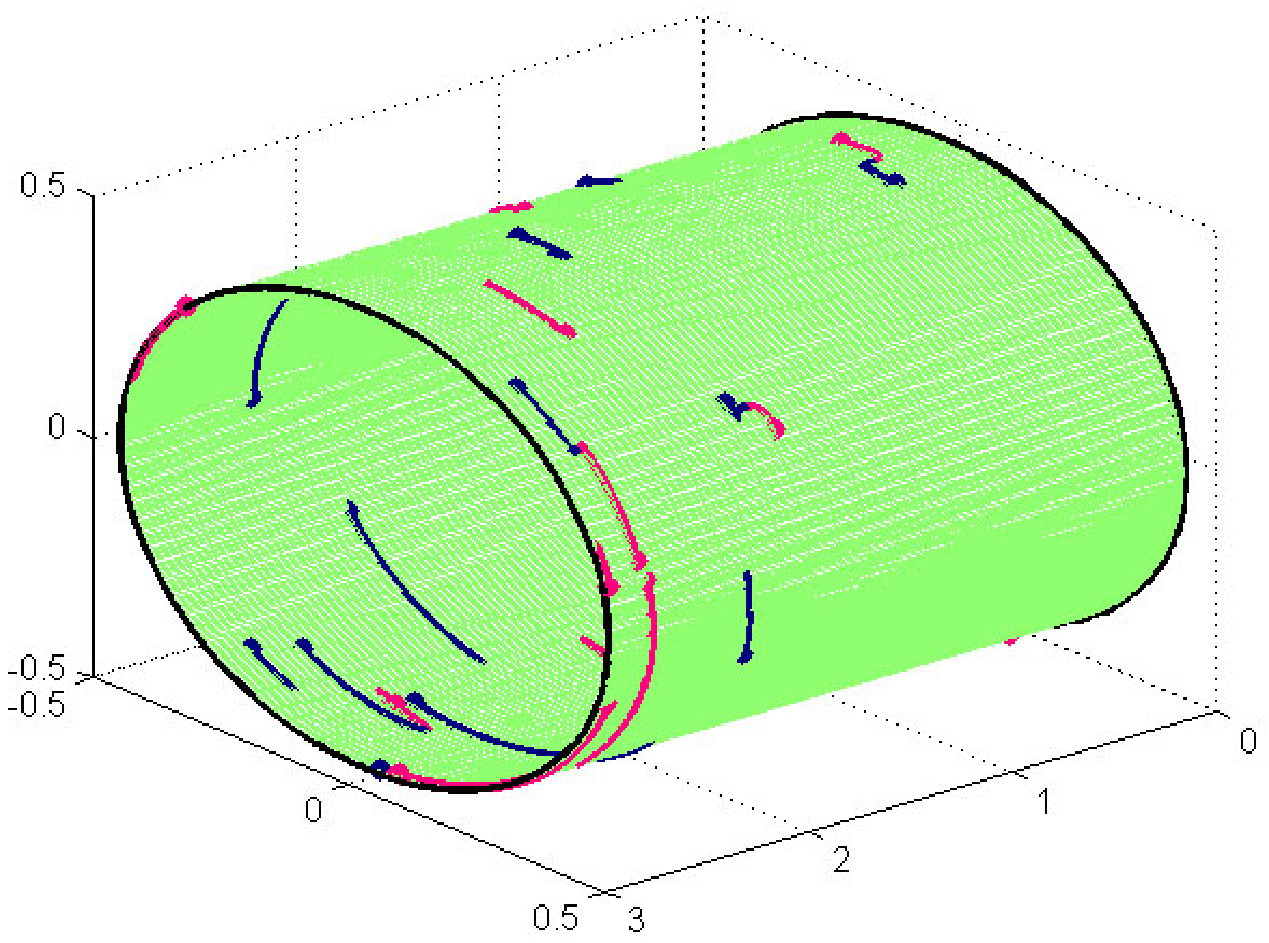}
\caption { Numerical simulation of the dynamics of 30 active dislocations, with strong elastic interactions. The dislocations are driven by a combination of chemical forces and turgor pressure in the circumferential direction, proportional to the sign of the Burgers vector. The temperature is zero and
$\frac{2\pi R G}{Ab^2h}=3$. The $x$ and $y$ coordinates of the dislocations were chosen randomly and uniformly along the axes, and $W/L=1$. The climb to glide mobility ratio is $\mu_c/\mu_g=10$. The red and blue points mark the starting positions of the $\pm b$ dislocations, and the red and blue lines correspond to their trajectories. The black circles mark the end of the cylinder. A snapshot of the simulation is taken after $t=0.07 \frac{W}{\mu_c G}$.
 \emph{A video showing the dynamics is attached.}
} \label {simulation1}
\end{figure}
\begin{figure}[b!]
\includegraphics[width=0.5\textwidth]{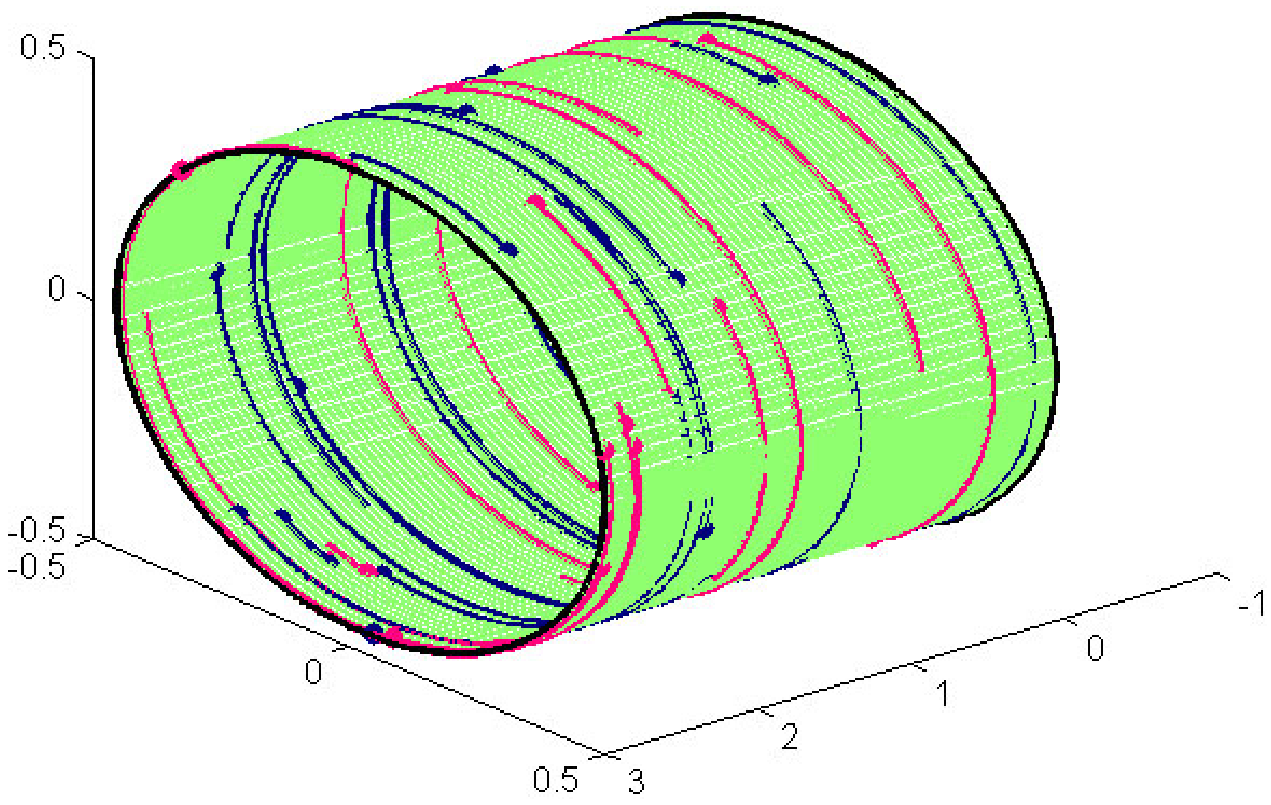}
\caption { Numerical simulation of the dynamics of 30 active dislocations, for the same conditions as for Fig. \ref{simulation1} but with a larger field $\frac{2\pi R G}{Ab^2h}=30$.
\emph{A video showing the dynamics is attached.}
} \label{simulation2}
\end{figure}

The dimensionless parameters for Fig 3 in the main text and Figs. 2 and 3 are described in Table 1.
\begin{tabular}{ c c c c c}
  \textbf{Param.} & \textbf{Fig 3 (main)}  & \textbf{Fig \ref{simulation1} (supp.)}  & \textbf{Fig \ref{simulation2} (supp.)}\\
  $\frac{G W}{h Ab^2}$ & 450 & 3 & 30\\
  $N_{in}$ &10,000 & 0 & 0 \\
  $\mu_c/\mu_g$ & 10 & 10 & 10\\
   $F_c/G$ & 1/3 & 100 & 10\\
  $d/W$ & 1 & 1 & 1 \\
  $T_s \mu_c \frac{G}{W} $ & 0.7 & 0.07 & 0.7
\end{tabular}

  \textbf{Table 1: Simulation parameters for Fig 3 in the main text and Figs. \ref{simulation1} and  \ref{simulation2} of the Supplementary Material.
 }

All relevant parameters for the simulations are given in the table using the relevant dimensionless ratios. In all simulations the number of active dislocations is 30, and the number of inactive ones, $N_{in}$, is varied. $W=2\pi R$ is the circumference and $d$ is length of the interval along the $x$ axis where the dislocations (active or inactive) were randomly chosen. All simulations use a ratio $d/W=1$, which is biologically relevant. $\frac{G W}{Ab^2h}$ indicates the relative strength of the driving field $G$ versus the elastic interactions, whose strength is given by the core energy $Ab^2h$. The fraction $\mu_c/\mu_g$ gives the mobility ratio between climb and glide, as explained above. $F_c$ determines the maximal elastic force created by the interactions of two dislocations, when their separation is of the order of the lattice spacing $b$. For Fig 3  in the main text, all parameters are chosen to be the biologically relevant ones. Figs 2 and 3 of the Supplementary Material have unrealistically large elastic interactions, and are used to illustrate the physics of the elastic interactions and their effects on the dynamics.

%Videos showing the time dynamics are attached, see S1.mpeg corresponding to Fig. 3 in the main text, S2.mpeg corresponding to Fig. 2 of the %Supplementary Material and S3.mpeg corresponding to Fig. 3 of the supplementary Material.

\vspace{10 mm}

\large
\textbf{Characterizing the quenched disorder}
\normalsize

As discussed in the manuscript, inactive dislocations interact with the active ones via long ranged elastic forces, thus creating a quenched disordered background. To estimate the typical magnitude of the disorder, we consider a typical E. coli cell with a length of 3 $\mu m$ and a circumference of 3 $\mu m$, and assume an average glycan strand length of 30 units, with a unit cell size of 2 $nm$. This implies approximately 10,000 inactive dislocations. To find the typical force associated with them, we consider 10,000 edge dislocations with a Burgers vector along the bacteria's long axis (with an equal number pointing in the positive and negative direction), and with a randomly chosen position on the cylinder. We then calculate the $\sigma_{xx}$ and $\sigma_{xy}$ components of the stress tensor at randomly chosen points. Fig. 4 shows the distribution of the resulting forces $F_x$ and $F_y$ exerted by this stress tensor on an edge dislocation. The standard deviation of these forces are found to be of the order of $10^{-6}$ \emph{dyne} for both
$\sigma_{xx}$ and $\sigma_{xy}$. Interestingly, this is the same order of magnitude as the force due to the turgor pressure and the biochemical forces, estimated in the main text. Hence, we expect that the interactions with the background of inactive dislocations can be significant, and could pin the motion of the active dislocations, as illustrated in Fig. 3 of the main text.
\begin{figure}
\includegraphics[width=8cm]{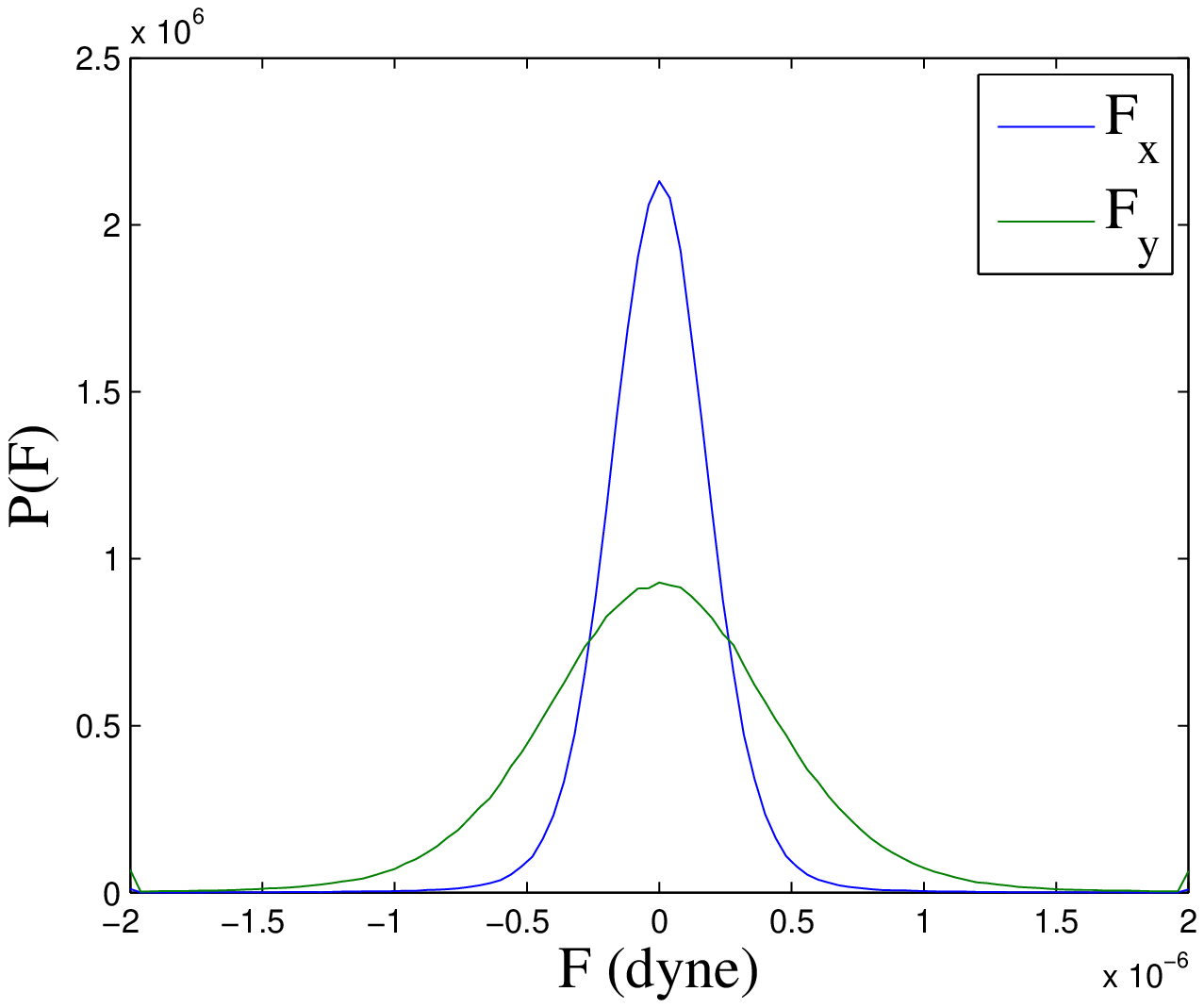}

\caption{Numerical determination of the force distribution due to the stress created by $N=10,000$
inactive dislocations randomly positioned on a cylinder of length
$L=3 \mu m$  and with a circumference of $W=3 \mu m$. All inactive
dislocations were taken as edge dislocation with a Burgers vector
along the bacterium's long axis, with an equal number of dislocations
for each sign of the Burgers vector. 100 disorder realizations were generated, and for each the total stress was evaluated
for 10,000 randomly chosen points, using Eqs. (\ref{sigma_xx_cylinder})
and (\ref{sigma_xy_cylinder}), with $Ab^2h=2 \cdot 10^-13$\emph{ erg}. The figure shows the distribution
of the $x$ and $y$ components of the forces exerted by this stress tensor on an edge dislocation with $\vec{b}=\hat{x}$, using Eqs. (\ref{force_x}) and (\ref{force_y}). \label{disorder_characterize}}
\end{figure}

%generated by noise_characterize_random_inactive.m

\vspace{10 mm}

\large
\textbf{Estimation of the biological parameters}
\normalsize

We estimate the relevant biological parameters for the case of \emph{E. coli}, to isolate the important physical processes.

\emph{Dislocation core energy:} The core energy associated with a single dislocation, $E_c$, arises from broken bonds and localized elastic disturbances, effects which we now estimate. According to Ref. \cite{yao}, the Young modulus of \emph{E. coli} is anisotropic,
with the elastic constant differing by roughly a factor of 2 between
the two symmetry axes (of the NAG-NAM chains and of the peptides \cite{young}).
The 3d Young modulus is of the order $y=2\cdot10^{7}N/m^{2}$, with similar values from Ref. \cite{shaevitz2}, which also discusses non-linear effects \cite{boulbitch}. Upon assuming the sacculus is approximately incompressible, we have a 3d Poisson
ratio $\nu=1/2$, and therefore $Y=2\mu(1+\nu)=3\mu$. In this limit we find $A \approx \frac{Y}{3\pi}$ in Eq. (3) in the main text.  The Burgers vector $b$ and the cell wall thickness $h$ should both be of order several \emph{nm} for gram negative bacteria like \emph{E. coli}. With $b \sim h \sim 2nm$, we find that the contribution of the elastic forces to the core energy is of the order of:
$E^{elastic}_{c}=hAb^{2}\sim hY\frac{b^{2}}{3\pi}\sim 2 \cdot 10^{-13}erg$. To create a dislocation, we must also take into account the energy of breaking the peptide bonds. Upon estimating this energy as 3 kcal/mole, we find an additional contribution of $E^{peptide}_c \sim 2 \cdot 10^{-13} erg$. Thus, we expect $E_c \sim 4 \cdot 10^{-13} erg$. This value is significantly higher than $k_BT$, which is approximately $4 \cdot 10^{-14} erg$ at 37 degrees Celsius.

%Another experimental observation is that the drift velocity in the
%$y$ axis is about $30nm/sec$. The diffusion coefficient is more
%difficult to estimate, but can be deduced from recent experiments \cite{garner}, giving a value of $D\sim20nm^{2}/sec$.
%This gives:

%$F=v/\mu=kTv/D=6\cdot10^{-7}dyn$.

\emph{Interactions between dislocations:}  The force due to a dislocation scales as $hAb^{2}\cdot1/distance=2\cdot10^{-13}erg/distance$. Let us find at which distance the interaction force between two dislocations would be comparable to that arising from the osmotic pressure. The latter seems to be of the order of an atmosphere in gram-negative bacteria \cite{shaevitz2}, which implies a stress tensor $\sigma \sim pR \sim 10^{5}Pa$, and a Peach-Kohler force $h\cdot b\cdot\sigma\sim10^{-6}$ \emph{dyne}. The resulting distance is of the order of a nanonmeter, suggesting that $G\gtrsim G^*$ as discussed above. If the main force on a dislocation arises from the turgor pressure, the Peach-Kohler force may help overcome interactions, unless cells are starved, enzymatic machinery is disabled, \emph{etc.} In healthy cells, with active glycan extension machinery, turgor pressure alone might be sufficient to drive the unbinding and subsequent motion of dislocations. There is, however, an additional complication: experiments suggest that the density of inactive dislocations far exceeds that of active dislocations, $n_{ac} \ll n_{in}$, since measured strand lengths are typically tens of repeat units \cite{Obermann, glycan}. This allows us to estimate the number of inactive dislocations as $N_{inactive}\approx10,000$, where we took an average strand length of 30, and a typical separation between adjacent strands of 2 \emph{nm}. As shown in the section ``Characterizing the disorder", for this scenario the typical net force due to this quenched potential is of the order of $10^{-6}$ \emph{dyne}, comparable to the ``driving force" $G$. Thus, we expect elastic interactions
with the inactive dislocations can be important. This effect is illustrated in Fig. 2 of the main text, showing that in this regime the interaction with the inactive dislocations can pin the motion of the active ones. For weaker driving forces, we expect that this effect will be even more pronounced, and could significantly reduce the growth rate. The relatively short glycan strand lengths measured in experiments can be reconciled with relatively long measured processivities in two way: Long glycan strands can be cut by enzymes after they have been inserted by the elongation machinery, as described by the term $\gamma_4$ in the rate equations. This could be potentially useful to the cell wall growth process, as a source of inactive dislocations which can later be rendered active. Alternatively, it could be that the elongation machinery occasionally `misses' an insertion of a NAM or NAG unit, thus creating a break in the glycan strand, without having the elongation machinery detach from the cell wall.

\end{document}